\documentstyle[art12]{article}

\begin{document}


\begin{titlepage}

\begin{flushright}
INP MSU Preprint-95-22/386\\
QMW-PH-95-22\\
hep-ph/95?????\\
\today
\end{flushright}

\begin{center}
{\Large\bf{
{On probing  the anomalous  magnetic moments of vector leptoquarks\\
in $ep$ collisions.}
}}
\end{center}
\bigskip
\begin{center}
{V. Ilyin, A. Pukhov, V. Savrin, A. Semenov, \\
{\small\it Institute of Nuclear Physics of Moscow State University,
        119899 Moscow, Russia}\\
 and \\
 W. von Schlippe \\
{\small\it Queen Mary \& Westfield College, London, England}
 }
\end{center}\bigskip

\vfill

\begin{abstract}
We study a possibility to measure the anomalous magnetic moment of vector
leptoquarks in reactions with single leptoquark production associated
with hard photon emission in $ep$ collisions.  For this purpose we propose
to use the {\it radiative amplitude zero} effect.  We find that an exact
radiative zero in the angular distribution of the emitted photon is present
only for pure Yang-Mills coupling of photons to leptoquarks.  As a result the
cross section is sensitive to the anomalous magnetic moment $\kappa$ of the
leptoquark. This effect offers a possibility to measure $\kappa$ with rather
high accuracy. For LEP+LHC we establish upper bounds in the plane leptoquark
mass--anomalous magnetic moment.
\end{abstract}

\vfill
\begin{center}
{(To be submitted to Physics Letters)}
\end{center}
\end{titlepage}

\section{Introduction.}

In this paper we continue the investigation of single leptoquark (LQ)
production with emission of a hard photon in electron-proton collisions,
\begin{equation}
                e^\pm +  p\rightarrow \gamma +LQ+X.
                                              \label{eq:ep-gammLQX}
\end{equation}
In \cite{paper1} we found that a radiative amplitude zero (RAZ) effect is
present for some types of leptoquarks.  Of particular interest is the case of
vector leptoquarks which have an exact radiative amplitude zero only in the
case of a Yang-Mills structure of coupling to the photon.  In this paper we
develop the application of this effect to the measurement of the anomalous
magnetic moment of the vector leptoquark. We present the numerical analysis
for the case of a possible LEP+LHC experiment \cite{LEP-LHC}.\footnote{At
HERA the cross sections for these processes are too small to be useful for
this purpose even for electroweak values of the LQ-fermion coupling constant
$\lambda=0.3$.}

\section{Anomalous magnetic moment}

The electromagnetic interaction of vector leptoquarks
is described by the following Lagrangian:
\begin{eqnarray}
   {\cal L}^V_\gamma = ieQ[A_\mu\Phi^{+\mu\nu}\Phi_\nu
    &-& A_\mu\Phi_\nu^+\Phi^{\mu\nu}+
        (1-\kappa) A_{\mu\nu}\Phi^{+\nu}\Phi^\mu] \nonumber \\
    &-& (eQ)^2[A^2(\Phi^+_\mu\Phi^\mu)-(A^\mu\Phi^+_\mu)(A^\nu\Phi_\nu)].
                                              \label{eq:vLQ-lagr}
\end{eqnarray}
Here $\Phi_{\mu\nu}=\partial_\nu\Phi_\mu-\partial_\mu\Phi_\nu$,
$A_{\mu\nu}=\partial_\nu A_\mu-\partial_\mu A_\nu$;
 $\Phi_\mu$  denotes the vector leptoquark; $e$ is  the elementary electric
charge; $Q$ is the charge of the leptoquark in units of $e$.

This Lagrangian contains a free dimensionless parameter $\kappa$, the {\it
anomalous magnetic moment} of the leptoquark.  The value $\kappa=0$
corresponds to the Yang-Mills structure of the interaction, while $\kappa=1$
corresponds to {\it minimal coupling}.  The measurement of this parameter
could help us in the understanding of the nature of this new massive vector
boson.

A value of $\kappa=0$ would indicate that the vector leptoquark belongs to
the family of gauge bosons. Such particles were introduced in grand unified
theories (GUT) (see the corresponding references in
\cite{LQ-models}\footnote{Here we are citing also some reviews where
leptoquarks were considered in the framework of different extensions of the
Standard Model.}).  In these models vector leptoquarks are represented by
some components of a unified gauge field that provides the interaction of all
matter fields, while the other components represent the known gauge bosons,
i.e. photons, gluons, $W$ and $Z$ bosons. Therefore in these theories the
Yang-Mills structure of the LQ-photon vertex is defined by the
selfinteraction of the unified gauge field.

On the other hand if it were found that
$\kappa=1$, then we should consider this leptoquark to be a new charged
matter field rather than a gauge field.  In this case the minimal coupling to
the electromagnetic field is also consistent with the principle of local
gauge invariance similarly to the electromagnetic interaction of leptons and
quarks.  However, in this case there arises a serious problem of how to
ensure the unitarity and renormalizability of the theory. Thus from a
theoretical point of view the deviation from the Yang-Mills structure is a
most interesting effect and we organize our numerical analysis in such a way
as to study its possible manifistations.

We can conclude that the measurement
of the leptoquark magnetic moment will arise as a fundamental problem when
these massive vector bosons are discovered. In the literature the possible
existence of vector leptoquarks with relatively small masses (hundreds of
GeV) is under discussion. Searches for leptoquarks have been carried out at
HERA and lower limits on their masses have been established at the level
$\sim 200$ GeV, depending on the leptoquark type \cite{HERA}.  Let us note
that some higher limits have been derived by indirect methods from low energy
experiments \cite{LowEn}, being stronger for the vector leptoquarks.  In
\cite{ee-LQ} the minimal coupling was investigated in leptoquark pair
production in $e^+e^-$ collisions. In \cite{Blum-Boos} the Lagrangian with
anomalous magnetic moment (and with anomalous quadrupole electric moment) was
considered in connection with probing these anomalous couplings in $e^+e^-$,
$\gamma e$ and $\gamma\gamma$ collisions. In the present paper we analyse the
potential of conceivable LEP+LHC experiments for the measurement of the
anomalous magnetic moment from reactions with single leptoquark production
associated with the emission of a hard photon.

\section{Squared matrix element and RAZ}

The hard subprocess underlying reaction (\ref{eq:ep-gammLQX}) is
\begin{equation}
              e^\pm + q\rightarrow \gamma +LQ,
                                              \label{eq:eq-gammLQ}
\end{equation}
where $q$ is a constituent quark of the proton. We consider the case when the
leptoquark interacts only with quarks of the first generation, i.e.
$q=(u,\bar u, d, \bar d)$.

{}From a phenomenological point of view the most general Lagrangian for
fermion-LQ vertices was proposed in \cite{BRW} where leptoquarks of all
possible quantum numbers were discussed.
Below we use the notation introduced for
leptoquarks in \cite{BRW} (see also the detailed table of leptoquark quantum
numbers in \cite{paper1}).

In \cite{paper1} we have obtained the analytical formulas for the squared
matrix element of hard subprocess (\ref{eq:eq-gammLQ})\footnote{The result in
the case of $U_3^{\pm 1}$ leptoquarks are greater by a factor of two.}:
\begin{equation}
|A|^2_{V}=\frac{e^2\lambda^2}{(\xi -1)^2}
\left[(Q_q \tau-Q_e \upsilon)^2 K_0\,+\,
Q (Q_q \tau-Q_e \upsilon) K_1 \kappa\;+\;
Q^2 K_2\kappa^2\right],
                                              \label{eq:vLQ-A2}
\end{equation}

$$
  K_0\equiv \frac{{\xi}^2+1-2\tau\upsilon}{\tau\upsilon},\qquad
  K_1\equiv \tau-\upsilon,\qquad
  K_2\equiv \frac{\tau\upsilon}{2}\; +\;
        \frac{\xi}{8} (\tau^2+\upsilon^2).$$
Here $Q_q$ and $Q_e = \mp 1$ are the quark, electron and positron charges,
respectively, in units of $e$; $Q=Q_q+Q_e$. We also use normalized
dimensionless Mandelstam variables ($M$ is the leptoquark mass):
$$
\xi\equiv \frac{\hat s}{M^2}\;=\;\frac{(p_e + p_q)^2}{M^2},
 \;\;\;\;\;\;
\tau\equiv \frac{t}{M^2}\;=\;\frac{(p_{\gamma} - p_e)^2}{M^2}, \;\;\;\;\;\;
\upsilon\equiv \frac{u}{M^2}\;=\;\frac{(p_{LQ} - p_e)^2}{M^2}.
$$

We see that the first two terms in (\ref{eq:vLQ-A2}) vanish at some value of
$\tau$ and $\upsilon$ due to the factor $(Q_q \tau-Q_e \upsilon)$.  This
means in particular that for $\kappa =0$ (Yang-Mills case) there is no photon
radiation in some direction depending on the electric charge of the
leptoquark ({\it radiative amplitude zero -- RAZ} effect).  In \cite{paper2}
we have investigated the RAZ effect in detail and arrived at some conclusions
about the possibility of observing the RAZ effect at HERA and LEP+LHC and to
use this effect for the determination of quantum numbers of the discovered
leptoquarks.

The fact that the $\kappa^2$ term has no RAZ factor $(Q_q \tau-Q_e \upsilon)$
makes the cross section sensitive to the value of the leptoquark anomalous
magnetic moment. In Fig. \ref{fig:sdifVRAZ} we show a typical angular
distribution with a clear dependence on $\kappa$ in the RAZ region. In the
next section we analyse numerically the possibility to get an upper bound on
the value of the anomalous magnetic moment from reaction
(\ref{eq:ep-gammLQX}). We show that the RAZ effect, if it is present, enables
one to establish much stronger upper bounds on the value of $\kappa$ then in
the cases without RAZ.

\section{Cross sections}

We have calculated the contributions of separate (quark) constituents to the
integrated cross sections by convoluting the differential cross sections of
the hard subprocesses with the corresponding parton distribution functions:
\begin{equation}
   \sigma(s)\;=\; \int^1_{x_{min}} dx\; q(x,Q^2)\, \int^1_{-1}
               d\cos{\vartheta_\gamma}
               \,\frac{d\hat\sigma(\hat s,\cos{\vartheta_\gamma},\kappa)}
               {d\cos{\vartheta_\gamma}} \,
               \,\Theta_{cuts}(E_{\gamma},\vartheta_\gamma).
                                               \label{eq:i-sigma-gammLQ}
\end{equation}
Here $\hat\sigma$ is the cross section of the corresponding subprocess
(\ref{eq:eq-gammLQ}); $s$ is the squared CMS energy of the electron-proton
system; the squared CMS energy of the hard subprocess is $\hat s=xs$; the
quark distribution function is denoted by $q(x,Q^2)$ and the 4-momentum
transfer scale is taken to be $Q^2 = \hat s$.  We denote the photon emission
angle by $\vartheta_\gamma$; the direction $\vartheta_\gamma=0$ is along the
proton beam.  The function $\Theta_{cuts}(E_{\gamma},\vartheta_\gamma)$
introduces the necessary kinematical cuts. The process under discussion is
infrared divergent, therefore we have to introduce a cut $E_\gamma >
E^0_\gamma >0$; for our numerical analysis we use $E^0_\gamma = 1$ GeV. Also
we introduce a cut on the photon emission angle, $\vartheta^{min}_\gamma <
\vartheta_\gamma <\vartheta^{max}_\gamma $, to exclude the unobservable
forward and backward cones and to ensure the optimal conditions for probing
$\kappa$ (see next section).  As the lower bound $x_{min}$ we use values
which will also ensure the optimal conditions for probing $\kappa$ .

For our calculations, both analytical
for the squared matrix elements and numerical for the integrands including
the convolution with parton distributions and $\Theta_{cuts}$, we have used
the CompHEP package \cite{CompHEP}. As Monte Carlo integrator and event
generator we have used the BASES/SPRING package \cite{BASES}.  For the parton
densities we used the parametrizations CTEQ2p \cite{CTEQ} and MRS-A
\cite{MRS} which take account of recent HERA data. Both parametrizations
gave the same results within calculation errors.

The numerical analysis of the corresponding cross sections is given for the
LEP+LHC collider with $\sqrt{s}=1740$ GeV, electron beam energy of $100$ GeV
and an integrated luminosity of $1\,\mbox{fb}^{-1}$.

Of course in reality the leptoquarks decay into two fermions, a lepton and a
quark. So we have to take account of complete sets of diagrams, including the
Standard Model background consisting of deep inelastic scattering associated
with hard photon emission. In \cite{paper2} we have calculated all
contributions and found that the Standard Model background was less than 1\%
of the signal.  This result was obtained for some set of kinematical cuts to
reduce the Standard Model background; the most effective cut was the cut on
the invariant mass of the lepton-quark system around the leptoquark mass.

\section{Probing the anomalous magnetic moment \label{k-analysis}}

In this section we analyse numerically the possibility of getting
an upper bound on $\kappa$ from experiment (\ref{eq:ep-gammLQX}).

One of the general restrictions on the LQ-fermion interaction is the
chirality of the lepton (see \cite{BRW} and references therein).  Therefore
we consider only either coupling with left-handed leptons or with
right-handed ones.  In both cases we
have used an electroweak value for this constant, $\lambda=0.3$.  For other
values of $\lambda$ numerical estimates can be obtained by rescaling our
results from the figures (the cross sections have $\lambda^2$ as a factor).
As to leptoquark masses we investigated the range
$200\,\mbox{GeV}<M<1\,\mbox{TeV}$.

Let us consider as a criterion for the value $\kappa\neq 0$ to be detectable
the condition that the number of events in this case has to be different from
those of the Yang-Mills case with $95\%\ CL$.  Therefore the following
relation must be satatisfied: $N(\kappa)-N(0)\,>\, 1.96\sqrt{N(0)}$.  Here
$N(\kappa)$ is the number of events corresponding to cross section
(\ref{eq:i-sigma-gammLQ}) with  luminosity $1\,\mbox{fb}^{-1}$.  It is
clear that the cross section is a quadratic function of $\kappa$. We found
from our numerical analysis that the minimum of this function lies near the
Yang-Mills value $\kappa=0$ and in the region under consideration its exact
position does not stronlgly depend on the leptoquark mass.  Therefore we
consider only the positive branch of $\kappa$. The analysis for negative
values gives practically the same results as for positive values: the
difference is no more than 20\%. Thus, we should keep in mind that if a
deviation from the Yang-Mills cross section is observed in the experiment it
will show evidence only for an absolute value of $\kappa$ but will not allow
us to determine its sign.  We denote by $\kappa_2$ the lowest positive value
of $\kappa$ for which the above criterion is satisfied.  It is clear that
larger values of $\kappa_2$ mean a lower sensitivity of the experiment to
$\kappa$; we have therefore chosen $\kappa_2$ as a
representative characteristic of our
analysis.

As we already noted the RAZ effect should increase the sensitivity to
$\kappa$.  Indeed, we are going to observe the contribution of the terms
proportional to $\kappa$ and to $\kappa^2$ in (\ref{eq:vLQ-A2}) with a value
of $\kappa$ as small as possible.  So regions, where the contribution of the
Yang-Mills term (the first term in (\ref{eq:vLQ-A2})) is smaller, are more
promising for our analysis.  It is clear that the region near the RAZ, where
$Q_q \tau=Q_e \upsilon$, has to be very sensitive to the value of $\kappa$.
We conclude that the strongest bound on the value of $\kappa$ can be obtained
for leptoquarks with RAZ, i.e. for $V_2$ and $U_3$ in the left-hand sector
and for $V_2$ and $\tilde U_1$ in the right-hand sector.  In Fig.
\ref{fig:k2theta} we show a typical dependence of parameter $\kappa_2$ on
the cut over the photon emission angle in the case of RAZ. We see from these
figures that an optimal cut could be given by
\begin{equation}
         10^\circ\; <\; \vartheta_\gamma\; <\; 90^\circ.
                                               \label{eq:thet-10-90}
\end{equation}

The term with the anomalous magnetic moment in Lagrangian (\ref{eq:vLQ-lagr})
contains derivatives of the electromagnetic field.  Therefore the
contribution of the $\kappa$ terms to the cross section is smaller for lower
energies of the emitted photon.  This point is important for our analysis
because the region near the threshold where the quasi-resonant peak
lies\footnote{We refer to \cite{paper2} where a typical distribution in $x$
is shown, with a sharp quasi-resonant peak near the threshold $x_0=M^2/s$
even with the cut $E_\gamma > 1$ GeV.} turns out
 to be insensitive to $\kappa$.  So to ensure a
maximal sensitivity to $\kappa$ we have to cut off the threshold region. On
the other hand, the introduction of too large a cut on $x$ could remove all
events. The implication is that we have to find the optimal value of
$x_{min}$.  In Fig. \ref{fig:d2xmin} we show $\kappa_2$ as a function of
$x_{min}$ for three values of the leptoquark mass and with the angular cuts
(\ref{eq:thet-10-90}) applied.
The dependence of $\kappa_2$ for large values of $x_{min}$ is rather smooth.
We found that the optimal values for this cut could be parametrized by the
following empirical formula
\begin{equation}
     x^{opt}_{min}\;=\;(M+150\,\mbox{\small GeV})^2/s.
                                                        \label{eq:xopt}
\end{equation}

Our final results are represented on the Fig. \ref{fig:d2m} where the
dependence of the $\kappa_2$ parameter is given as a function of the
leptoquark mass, on the left-hand side in the RAZ case and on the right-hand
side for other leptoquarks.  Points above the curves can be measured with
$95\%\ CL$.  Here we have used the cuts (\ref{eq:thet-10-90}) and
(\ref{eq:xopt}). In Table \ref{tab:kapupmassRAZ} we give the upper bounds on
the leptoquark masses for two values of the anomalous magnetic moment,
$\kappa=1$ and $\kappa=3$ in the RAZ cases. For lower masses the
corresponding values of $\kappa$ can be measured with $95\%\ CL$. Let us
stress again that only an absolute value of $\kappa$ is discussed here.

In this paper
we have carried out the RAZ analysis only for reactions with electron decay
modes of the leptoquarks.  So only some isospin components of the leptoquarks
were considered.  The analysis for the complementary isospin components has
to be based on reactions with neutrino decay modes. These channels could be
added also in the total statistics to establish stronger bounds on $\kappa$.
The corresponding analysis will be presented elsewhere.

\section{Conclusions}

The proposed analysis of cross sections of vector leptoquark production in
$ep$ collisions shows a possibility to measure the leptoquark anomalous
magnetic moment $\kappa$. The corresponding experiments will be more
sensitive for those leptoquarks which reveal the RAZ effect.

The value $\kappa=1$ (minimal coupling) can be measured at LEP+LHC up to
$M=400\,\mbox{GeV}$ for $V_2^{-{1\over2}}$, and up to $M=600\,\mbox{GeV}$ for
$U_3^{-1}$ and $\tilde U_1$ types of leptoquarks. In the case of $U_3$ and
$\tilde U_1$ leptoquarks, the positron-proton channel is more sensitive than
the electron-proton channel.

The anomalous magnetic moment of leptoquarks for which there is no RAZ effect
can be measured only on the level of several units or even $\kappa \sim 10$.

\section*{Acknowlegements}

This work was partly supported by The Royal Society of London as a joint
project between Queen Mary \& Westfield College (London) and Institute of
Nuclear Physics of Moscow State University, and by INTAS (project 93-1180).
V.I. and V.S. wish to thank QMW for hospitality and for the possibility to
work during their visit to London.  WvS wishes to thank INP-MSU for
hospitality. The work of V.I., A.P. and A.S. was partly supported by Grants
M9B000 and M9B300 from the International Science Foundation.

\clearpage
\newpage

\section*{Table}

\begin{table}[h]
\begin{center}
\begin{tabular}{|c|c|cccc|}
\hline
\multicolumn{2}{|c|}{\raisebox{0ex}[3ex][1ex]{$\ell q$ channel}}
          & $e^+\bar d$ & $e^-\bar u$ & $e^- d$ & $e^+ u$ \\
\hline
\multicolumn{2}{|c|}{\raisebox{0ex}[3ex][1ex]
     {left-handed lepton}}
   & $V_2^{-{1\over 2}}$ & $U_3^{-1}$ & $V_2^{-{1\over 2}}$ & $U_3^{-1}$ \\
\multicolumn{2}{|c|}{right-handed lepton}
& $V_2^{-{1\over 2}}$ & $\tilde U_1$ & $V_2^{-{1\over 2}}$ & $\tilde U_1$ \\
\hline
 & $\kappa=1$ &  310 & 370 & 390 & 590 \\
$\mbox{max}\,M\,\mbox{\small [GeV]}$ & $\kappa=3$
              & 530 & 590 & 660 & 890 \\
\hline
\end{tabular}
\end{center}
\caption{Upper bounds on the LQ mass, for lower masses the corresponding
values $\kappa=1$ and $\kappa=3$ can be measured with $95\% CL$.}
\label{tab:kapupmassRAZ}
\end{table}

\vskip 3cm

\section*{Figure captions}

\begin{figure}[h]

\caption{RAZ effect for
       $V^{-{1\over 2}}_2$ with $M=300\,\mbox{GeV}$.
       Here $x_{min}=0.04$.
       For $\kappa=0\,(1)$ the cross section equals
             $\sigma=2.41\,(2.49)\,\mbox{pb}$.
\label{fig:sdifVRAZ} }

\caption{
Left-hand side: $\kappa_2$ {\it vs} $\vartheta_\gamma^{min}$
                with $\vartheta_\gamma^{max}=90^\circ$;
Right-hand side: $\kappa_2$ {\it vs} $\vartheta_\gamma^{max}$
                with $\vartheta_\gamma^{min}=10^\circ$;
different curves correspond to:
  1) $M=300\,\mbox{GeV}$, $x_{min}=0.067$;
  2) $M=400\,\mbox{GeV}$, $x_{min}=0.1$;
  3) $M=500\,\mbox{GeV}$, $x_{min}=0.14$.
\label{fig:k2theta} }

\caption{$\kappa_2$ {\it vs} $x_{min}/x_0$.
  Here $\vartheta_\gamma^{min}=10^\circ$,
  $\vartheta_\gamma^{max}=90^\circ$;
  different curves correspond to: 1) $M=300\,\mbox{GeV}$;
  2) $M=400\,\mbox{GeV}$ and 3) $M=500\,\mbox{GeV}$.
\label{fig:d2xmin} }

\caption{$\kappa_2$ {\it vs} LQ mass.
  Here $\vartheta_\gamma^{min}=10^\circ$,
  $\vartheta_\gamma^{max}=90^\circ$; $x_{min}=x^{opt}_{min}$.
  On the left-hand side processes with RAZ are presented,
  on the right-hand side RAZ effect is absent.
  \label{fig:d2m} }
\end{figure}

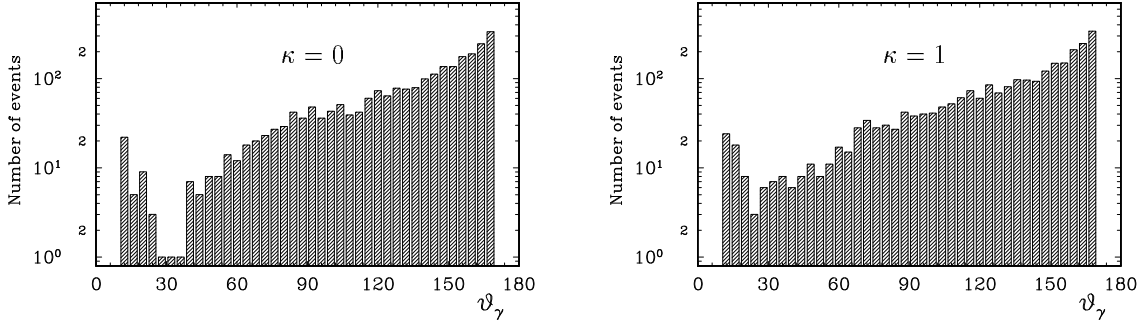
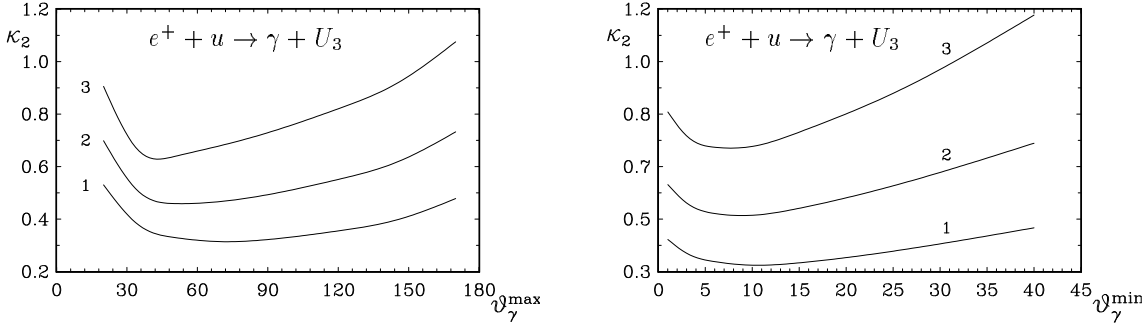
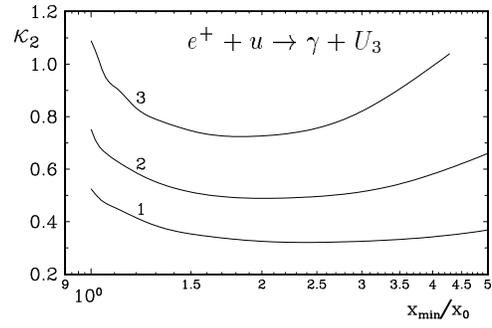
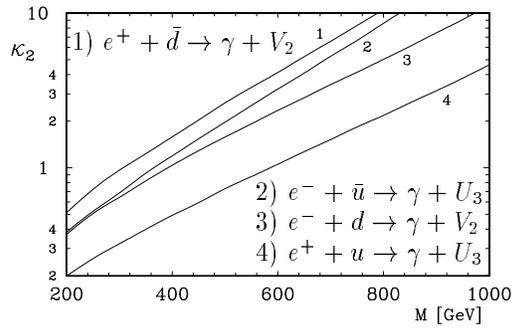
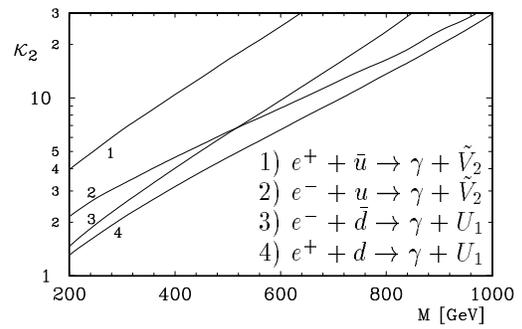
\end{document}